\documentstyle[preprint,aps]{revtex}

\begin{document}
\newcommand{\beq}{\begin{equation}}
\newcommand{\eeq}{\end{equation}}
\newcommand{\beqa}{\begin{eqnarray}}
\newcommand{\eeqa}{\end{eqnarray}}
\newcommand{\sr}{\sqrt}
\newcommand{\fr}{\frac}
\newcommand{\mn}{\mu \nu}
\newcommand{\G}{\Gamma}

\draft
\preprint{ INJE-TP-99-10}
\title{Comment on ``Metric Fluctuations in Brane Worlds"
}
\author{Y.S. Myung\footnote{E-mail address:
ysmyung@physics.inje.ac.kr} and Gungwon Kang\footnote{E-mail address:
kang@physics.inje.ac.kr}}
\address{
Department of Physics, Inje University, 
Kimhae 621-749, Korea}
\maketitle
\begin{abstract}

Recently, Ivanov and Volovich (hep-th/9912242) claimed 
that the perturbation of $h_{\mu\nu}$
with nonvanishing transverse components $h_{5\mu}$ is not localized 
on the brane because 
$h_{\mu\nu}$ depends on the fifth coordinate $z$ linearly. 
Consequently, it may indicate that
the effective theory is unstable. However, we point out that such 
linear dependence on $z$ can be {\it gauged away}. Hence the solution does 
not belong to the physical one. Therefore, even if one includes 
$h_{5\mu}$, Randall and Sundrum's argument for the localized gravity
on the brane remains correct.   

\end{abstract}
\bigskip

\newpage
Recently, Ivanov and Volovich~\cite{IV} have found a solution for
metric fluctuations including 
nonvanishing transverse components given by 
\beqa
h_{55} &=& 0, \nonumber  \\
h_{5\mu} &=& c_{\mu} e^{ip\cdot x}, \nonumber  \\
h_{\mu\nu} &=& \Big[ c_{\mu\nu} +i(c_{\mu}p_{\nu}+c_{\nu}p_{\mu})z \Big] 
               e^{ip\cdot x}
\label{hmn}
\end{eqnarray} 
with $p_{\mu}p^{\mu} =0$, $c_{\mu}p^{\mu} =0$, $\eta^{\mu\nu}c_{\mu\nu}
=0$, and $c_{\mu\nu}p^{\nu}=0$. Here we choose the signature $\eta_{\mu\nu}
= {\rm diag}(+,-,-,-)$. $c_{\mu}$, $c_{\mu\nu}$, and $p_{\mu}$ are
constants.

In the paper, the de Donder gauge for $h_{MN} (M,N =\mu , 5)$ were
chosen as
\beq
\partial_M h^{MN} =0 \qquad \qquad {\rm with} \qquad h^P_P=0.
\label{gauge}
\eeq
The above gauge cannot eliminate all gauge degrees of freedom. 
Even after choosing the gauge conditions Eq.~(\ref{gauge}), actually
there remains a sort of residual gauge degrees of freedom 
as follows~\cite{Wein};
\beq
h'_{MN} = h_{MN} - \partial_M \xi_N - \partial_N \xi_M, 
\eeq
which are obtained from the coordinate transformations $x^{M} \rightarrow 
x^M + \xi^M$. Notice that $h'_{MN}$ also satisfy the de Donder gauge
Eq.~(\ref{gauge}) provided that 
\beq
\partial_M \xi^M =0, \qquad \qquad  \Box \xi^M =0.
\label{gauge2}
\eeq
In other words, both $h_{MN}$ and $h'_{MN}$ describe the same physical 
situation. 

Now suppose that we choose
\beq
\xi_M = \varepsilon_M e^{ip\cdot x}.  
\eeq
$\xi_M$ satisfies Eq.~(\ref{gauge2}) if $\varepsilon_M$ is a function 
of the fifth coordinate $x^5=z$ only, $\varepsilon_5=0$, 
and $\varepsilon_{\mu} p^{\mu} =0$. Then,
\beqa
h'_{55} &=& h_{55}, \qquad \qquad h'_{5\mu} = h_{5\mu} 
         - (\partial_z \varepsilon_{\mu}) e^{ip\cdot x}, \nonumber \\ 
h'_{\mu\nu} &=& h_{\mu\nu} - i(\varepsilon_{\mu}p_{\nu}
              + \varepsilon_{\nu}p_{\mu}) e^{ip\cdot x}.
\label{hprime}
\eeqa
Thus, if $h_{MN}$ were those in Eq.~(\ref{hmn}) and  
$\varepsilon_{\mu} = c_{\mu}z$,
then one finds 
\beqa
h'_{55} &=& h'_{5\mu} =0, \nonumber  \\
h'_{\mu\nu} &=& c_{\mu\nu}e^{ip\cdot x}. 
\eeqa
Therefore, we see that the z-dependent term in Eq.~(\ref{hmn})
indeed belongs to a gauge degree of freedom. 
In addition, our analysis shows that, in order to have nonvanishing
transverse components up to gauge freedom, one at least needs to
assume that $c_{\mu}$ is not constant, but $z$-dependent in 
Ref.~\cite{IV}. 


\end{document}